%% file: CLAS_L1405_spinparity6.tex
\RequirePackage{lineno} 

\documentclass[aps,prl,superscriptaddress,twocolumn,nofootinbib,amsmath,amssymb,showpacs,showkeys]{revtex4-1}

\usepackage{pifont}
\usepackage{graphicx}
\usepackage{color}
\usepackage{amsmath}
\usepackage{amssymb}
\usepackage[caption=false]{subfig}
\usepackage{multirow}

\newcommand{\proton}{\ensuremath{p}}

\newcommand{\pim}{\ensuremath{\pi^{-}}}
\newcommand{\kp}{\ensuremath{K^{+}}}

\newcommand{\pizero}{\ensuremath{\pi^{0}}}
\newcommand{\SigmaPlus}{\ensuremath{\Sigma^{+}}}

\newcommand{\LambdaOne}{\ensuremath{\Lambda(1405)}}
\newcommand{\LambdaTwo}{\ensuremath{\Lambda(1520)}}

\newcommand{\gevcc}{\ensuremath{\mathrm{GeV}/c^{2}}} 

\newcommand{\costhetakp}{\ensuremath{\cos \theta_{\kp}^{\mathrm{c.m.}}}}

\newcommand{\norm}[1]{|#1|}

\newcommand{\smallt}[2]{\ensuremath{t_{#1}^{#2}}}

\begin{document}

\title{Spin and parity measurement of the $\mathbf{\Lambda(1405)}$ baryon}

\input{authors}

\date{\today}

\begin{abstract}
A determination of the spin and parity of the
\LambdaOne{} is presented using photoproduction data from the CLAS
detector at Jefferson Lab. The reaction $\gamma + \proton \to \kp +
\LambdaOne$ is analyzed in the decay channel $\LambdaOne
\to \SigmaPlus + \pim$, where the decay distribution to $\SigmaPlus
\pim$ and the variation of the \SigmaPlus{} polarization
with respect to the \LambdaOne{} polarization direction determines the
parity.  The \LambdaOne{} is produced, in the energy range $2.55 < W <
2.85$~GeV and for $0.6 < \costhetakp < 0.9$, with polarization $P =
0.45 \pm 0.02 (\text{stat}) \pm 0.07 (\text{syst})$.  The analysis shows that the
decays are in $S$ wave, with the \SigmaPlus{} polarized such that the
\LambdaOne{} has spin-parity $J^{P} = \frac{1}{2}^{-}$, as expected by most
theories.
\end{abstract}

\maketitle


The \LambdaOne{} has long been a peculiar state in the spectrum of
excited hyperons. Lying just below the $N \overline{K}$ threshold,
there is no universal agreement on the nature of this state.  
The constituent quark model for $P$-wave baryons~\cite{Isgur-Karl_PRD18,*Capstick-Isgur,*Capstick:2000qj},
which has had success in describing the nonstrange low-mass baryons,
has difficulty in computing the correct mass. More recently,
the chiral unitary
approach~\cite{Oset-Ramos,*Oller:2000fj,*Hyodo:2011ur,*Ikeda:2012au}
describes the \LambdaOne{} as a dynamically generated
state of two overlapping isospin-zero poles in
the rescattering of the octet meson and baryon states that couple to
it. Another theory describes the state as a quasibound state
of $N \overline{K} $ embedded in a $\Sigma \pi$
continuum~\cite{Akaishi,*Akaishi:2010wt}.

In all of the above theories, a crucial assumption is that the
\LambdaOne{} has spin-parity $J^{P} = \frac{1}{2}^{-}$. This assumption is
somewhat justified by the proximity of the \LambdaOne{} mass to the
$\Sigma \pi$ and $N \overline{K}$ thresholds and also by
previous experimental analyses which saw a rapid fall in intensity of
the \LambdaOne{} line shape as it crossed the $N \overline{K}$
threshold~\cite{Hemingway,Thomas}. Only a state that couples to
$\Sigma \pi$ and $N \overline{K}$ in $S$ wave would show such
behavior, and currently, this is, in the words of Dalitz in the
1998 PDG review article, the ``sole direct evidence that $J^{P} =
\frac{1}{2}^{-}$''~\cite{Dalitz:PDG}. 
There is at least one
model~\cite{Kittel-Farrar1,*Kittel-Farrar2} that postulates the
\LambdaOne{} to be a $\frac{1}{2}^{+}$ state with $P$-wave
coupling to the $\Sigma \pi$ final state.

While many in the field would not doubt the assertion that the 
\LambdaOne has $J^P=\frac{1}{2}^{-}$
it is, nevertheless, important to have experimental confirmation.
Previous experiments~\cite{Engler,Thomas,Hemingway}
showed that the spin was
consistent with $\frac{1}{2}$, but insufficient statistics and lack of
polarization of the \LambdaOne{} 
made the parity determination impossible.  Recently, the results of a
photoproduction experiment by the CLAS Collaboration at the Thomas
Jefferson National Accelerator Facility (Jefferson Lab) were
presented~\cite{lineshapepaper,crosssectionpaper}. 
The exclusive reaction $\gamma +
\proton \to \kp + \LambdaOne$ was produced using an unpolarized beam 
and target and analyzed for the three $\Sigma \pi$ decay channels.
A rapid falloff of the line shapes was seen near the
$N\overline{K}$ threshold,
and fits were made to the line shapes assuming an $S$-wave
coupling to the $\Sigma \pi$ and $N \overline{K}$ final
states~\cite{lineshapepaper, Schumacher2013}. 


Consider an excited hyperon $Y^\ast$ of
spin-parity $J^{P}$ that decays strongly into $Y \pi$, where $Y$ is a
ground state hyperon.  The $Y \pi$ angular distribution is determined
solely by $J$ and not $P$. This can be shown by
using the statistical tensors given by \citet{ByersFenster}, where the
$Y^\ast$ spin density matrix is parametrized by parameters
$\smallt{L}{M}$ with $0 \leq L \leq 2J$ and $|M| \leq L$. For a
state that decays strongly, the decay distribution to $Y \pi$ depends
only on a subset of these parameters with $L$ even. Furthermore, the
$\smallt{L}{M}$ with $M$ odd vanish if one picks the spin quantization
axis, $\hat{z}$, out of the production plane.  Our coordinate system
is set up so that in the $Y^\ast$ rest frame and in the center-of-mass
(c.m.) frame for $\gamma + \proton \to K^{+} +
\LambdaOne$, the incoming photon direction is in the negative $y$ direction.  
The direction out of the production plane is defined by
$\hat{z} = \vec{p}_{\gamma} \times \vec{p}_{\kp} / \left|
\vec{p}_{\gamma}\times \vec{p}_{\kp} \right|$, where $\vec{p}_{\gamma}$ and
$\vec{p}_{\kp}$ are the momenta of the incoming photon and outgoing
\kp, respectively. For the case of spin $\frac{1}{2}$, one finds that there are no
remaining degrees of freedom in the decay distribution, so it is
always isotropic.  For spin $\frac{3}{2}$, the decay distribution is given by
\begin{align}
  I(\theta_{Y}) &\propto 1 + \frac{3(1-2p)}{2p + 1}
  \cos^{2}\theta_{Y}, \label{eq:distThreeHalf}
\end{align}
where $\theta_{Y}$ is the polar angle of the decay direction of $Y$ in
the $Y^\ast$ rest frame.  The parameter $p$ describes the $Y^{\ast}$
fraction with spin projections along the $z$ axis with $\pm \frac{3}{2}$ and
not $\pm \frac{1}{2}$ and is related to \smallt{2}{0}.  In general, a higher
degree of complexity in the decay distribution signals a higher
minimum spin that the state may possess, but an arbitrarily high spin
state may still mimic the simpler behavior of a lower spin state. In
the limit of completely unpolarized $Y^*$ production, the decay
distribution will be isotropic, and no information on the spin of the
state is obtainable. Conversely, this would mean that there is no
positive evidence that a state has spin $\frac{1}{2}$, since it could just be a
higher spin state that was produced unpolarized. Therefore, a fit to
the decay distribution that is consistent with isotropy is the best
possible evidence of spin $\frac{1}{2}$, but in general, this does not rule out
higher spins.

\begin{figure}[h!t!p!b!]
    \includegraphics[width=1.0\linewidth]{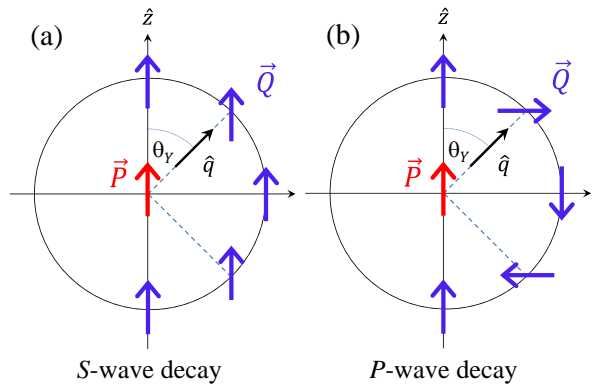}
    \vspace{-1.5cm} 

    \caption{(Color online) Polarization transfer from $Y^{\ast}$ to
    $Y$ in the decay $Y^{\ast} \to Y + \pi$, where $Y^{\ast}$ has
    spin $\frac{1}{2}$. The red arrow shows the polarization $P$ of the
    $Y^{\ast}$ taken to be in the $z$ direction, while the blue arrows
    show the polarization $\vec{Q}$ of $Y$ depending on the decay
    angle $\theta_{Y}$ around the $z$ axis.  (a) is for odd parity;
    (b) is for even parity.}  \label{fig:poltransfer}
\end{figure}

With the state's spin determined, the polarization that is transferred
from $Y^{\ast}$ to $Y$, called $\vec{Q}$ in
Fig.~\ref{fig:poltransfer}, is related to the odd-$L$ \smallt{L}{M}{}.
One finds that the longitudinal polarization of $Y$ along $\hat{q}$,
the decay direction in the $Y^\ast$ rest frame, will be independent of
the parity of $Y^{\ast}$, while the transverse polarization will
change sign depending on the parity. Since the parity of the $Y^{\ast}$ is
determined by the orbital angular momentum $L$ of the decay $Y^{\ast}
\to Y \pi$, and since a measurement of $\vec{Q}$ determines 
$L$, the parity can be found.

A schematic of two polarization transfer scenarios depending on the
parity of the $Y^{\ast}$ is shown in Fig.~\ref{fig:poltransfer}.  In the
case of spin $\frac{1}{2}$, the use of an unpolarized beam and target restricts
the polarization of the $Y^{\ast}$ to be in the direction out of the
production plane specified by $\vec{P} = P\hat{z}$ in
Fig.~\ref{fig:poltransfer}.  It can be shown that for an $S$-wave
decay ($J^{P}=\frac{1}{2}^{-}$) of $Y^{\ast}\rightarrow Y \pi$, $\vec{Q}$ is
independent of $\theta_Y$ and is given as $\vec{Q} = \vec{P}$; that
is, it retains the same polarization as that for the $Y^{\ast}$, so that
$Q_{z} = P$. For a $P$-wave decay ($J^{P} = \frac{1}{2}^{+}$), it is given by
\begin{align}
  \vec{Q} &= -\vec{P} + 2 \left( \vec{P} \cdot \hat{q} \right) \hat{q},
  \label{eq:pwavePol}
\end{align}
so that although the magnitude is unchanged, the direction depends on
$\hat{q}$.  If a component measurement is made of the polarization
along the original $Y^{\ast}$ polarization direction, then $Q_{z} =
P(2 \cos^{2} \theta_{Y} - 1)$ so that at $\cos \theta_{Y} = 0$,
$Q_{z}$ must have the opposite sign compared to $\cos \theta_{Y} = \pm
1$.

In all cases, the polarization of a ground state hyperon can be
measured by the weak decay asymmetry in its decay distribution into a
nucleon and pion.  Therefore, by measuring the polarization of the $Y$
for different decay directions in the $Y^{\ast}$ rest frame, we can
deduce the parity of the $Y^{\ast}$. This method requires that the
original $Y^{\ast}$ be produced polarized, but beyond that, there are
no further assumptions necessary to uniquely determine the spin and
parity.


The setup of the CLAS experimental run g11a used in this analysis has
been explained in Ref.~\cite{lineshapepaper}, and further details are
also available~\cite{Moriya-thesis}. 
Based on the results of our previous analyses~\cite{lineshapepaper,
  crosssectionpaper}, we select kinematic ranges where the
\LambdaOne{} is the dominant contribution in the $\Sigma \pi$ mass
range of interest. The nine bins of energy and angle we select have
c.m. energies $W$ centered at $2.6, 2.7$, and $2.8$ GeV, and for each energy bin,
the three forwardmost kaon angle bins were used.
The detected particles were $K^+$, $p$, and $\pi^-$, with a kinematic
fit applied to select events with a missing $\pi^0$.

\begin{figure}[h!t!p!b!]
  \subfloat{\includegraphics[width=\linewidth]{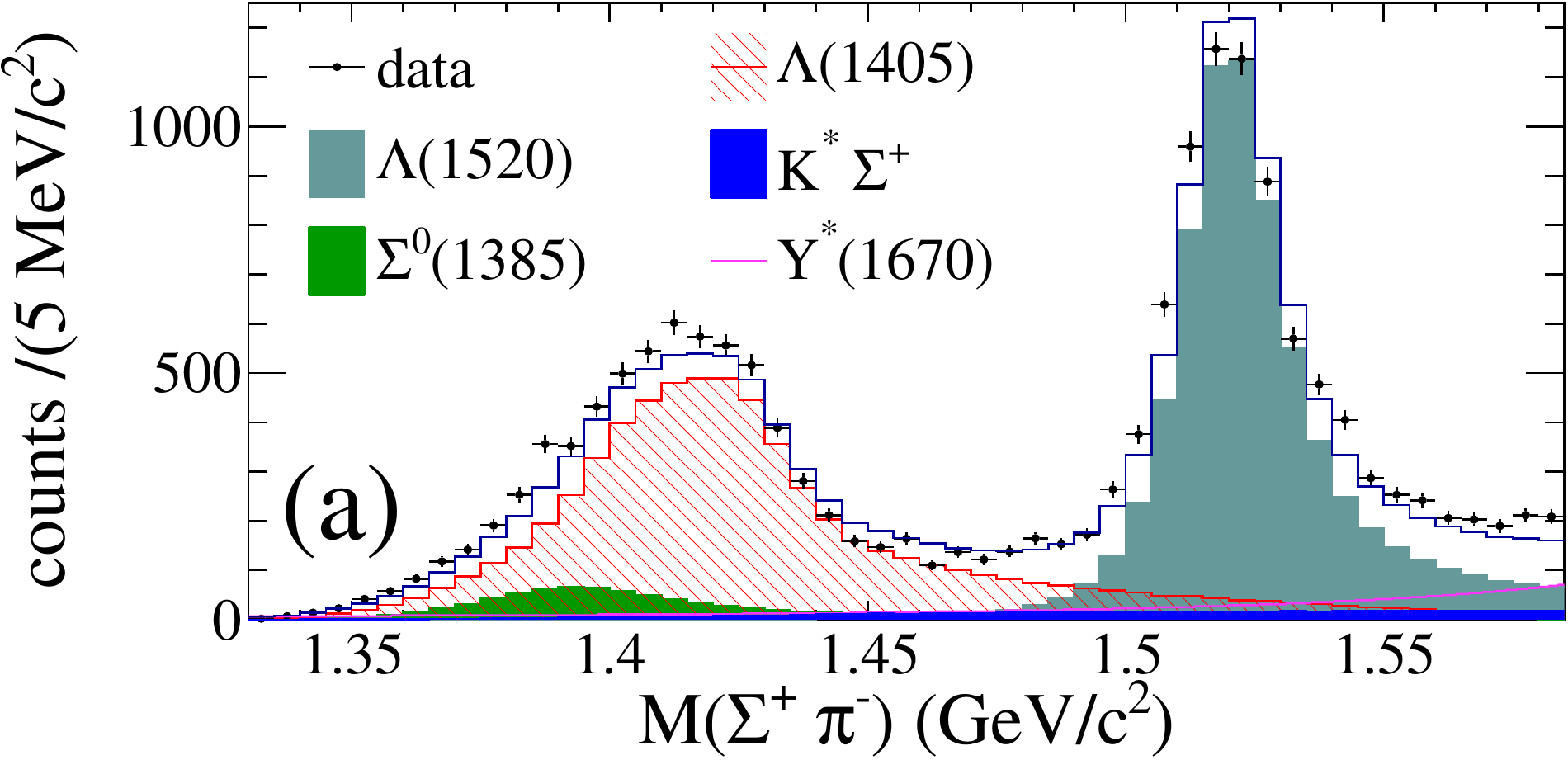}} \\
  \subfloat{\includegraphics[width=\linewidth]{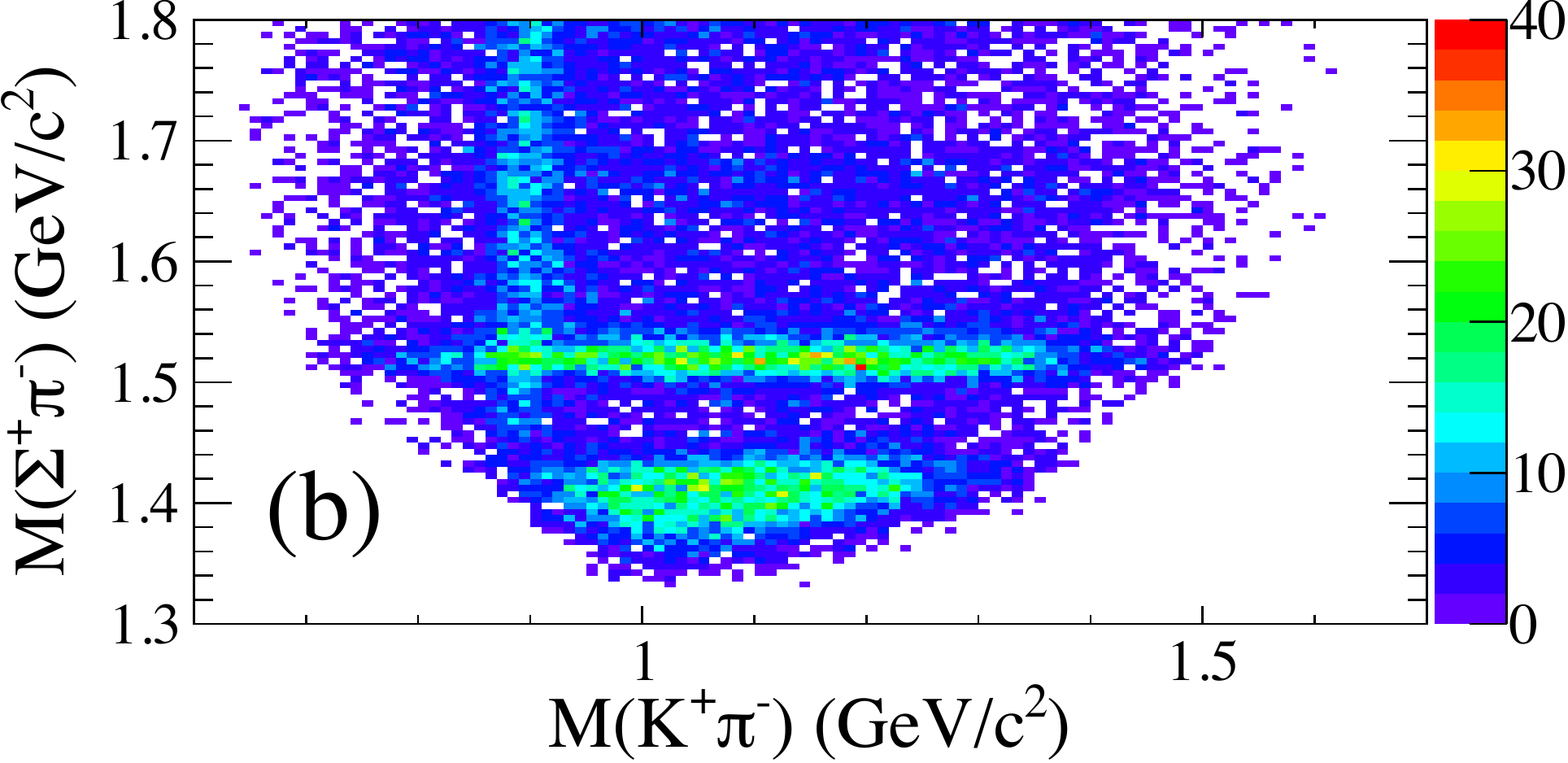}}

  \caption{(Color online) (a) $M(\SigmaPlus \pim)$ distribution summed
  over the range of $W$ and kaon angle used in this analysis. The
  \LambdaOne{} (hatched red histogram) and \LambdaTwo{} (filled cyan
  histogram) are seen.  The estimate for the main background due to
  the $\Sigma^0(1385)$ is superimposed (filled green area) near the
  bottom. Contributions from the $K^{\ast 0} \SigmaPlus$ (filled blue
  histogram) and $Y^{\ast}(1670)$ background (solid magenta curve) are
  also shown but contribute very little. The fit total is shown as the
  open blue histogram. (b) $M(\SigmaPlus \pim)$ versus
  $M(\kp \pim)$ for the range of $W$ and kaon angles used in
  the analysis.  The vertical band is due to $K^*$ production, while
  the horizontal bands show the \LambdaOne{} below and the
  \LambdaTwo{} above.  } \label{fig:data}
\end{figure}

Figure~\ref{fig:data}(a) shows the $\Sigma ^+\pi^-$ invariant mass
$M(\SigmaPlus \pim)$ for these nine bins combined, where the
\LambdaOne{} and \LambdaTwo{} are seen, along with background
contributions from $K^+ \Sigma^{0}(1385)$ and $K^{\ast 0}
\SigmaPlus$ production. In the present analysis the backgrounds could 
not be removed event by event.  Events were selected based on the 
$\SigmaPlus \pim$ invariant mass range of $1.30$--$1.45 ~\gevcc$,
where the spectrum is dominated by the
\LambdaOne.
Backgrounds due to non-\LambdaOne{} production were estimated in
previous works, where beside the channels listed above, a 
$Y^{\ast}(1670)$ background was used to parametrize the higher-mass data. We
estimate  these backgrounds as
approximately $16\%$ total, mostly from the $\Sigma^{0}(1385)$.

Figure~\ref{fig:data}(b) shows a Dalitz-like plot of the
$\SigmaPlus\pim$ invariant mass $M(\SigmaPlus\pim)$ versus the mass
$M(\kp\pim)$, where a slight overlap of the $K^{\ast 0} \SigmaPlus$
events (vertical band) with the \LambdaOne{} and \LambdaTwo{} events
(horizontal bands) is seen. Due to the kinematics of the reaction, the
overlap is not significant and has little influence in these
kinematic bins.

\begin{figure}[h!t!p!b!]
  \subfloat{\includegraphics[width=\linewidth]{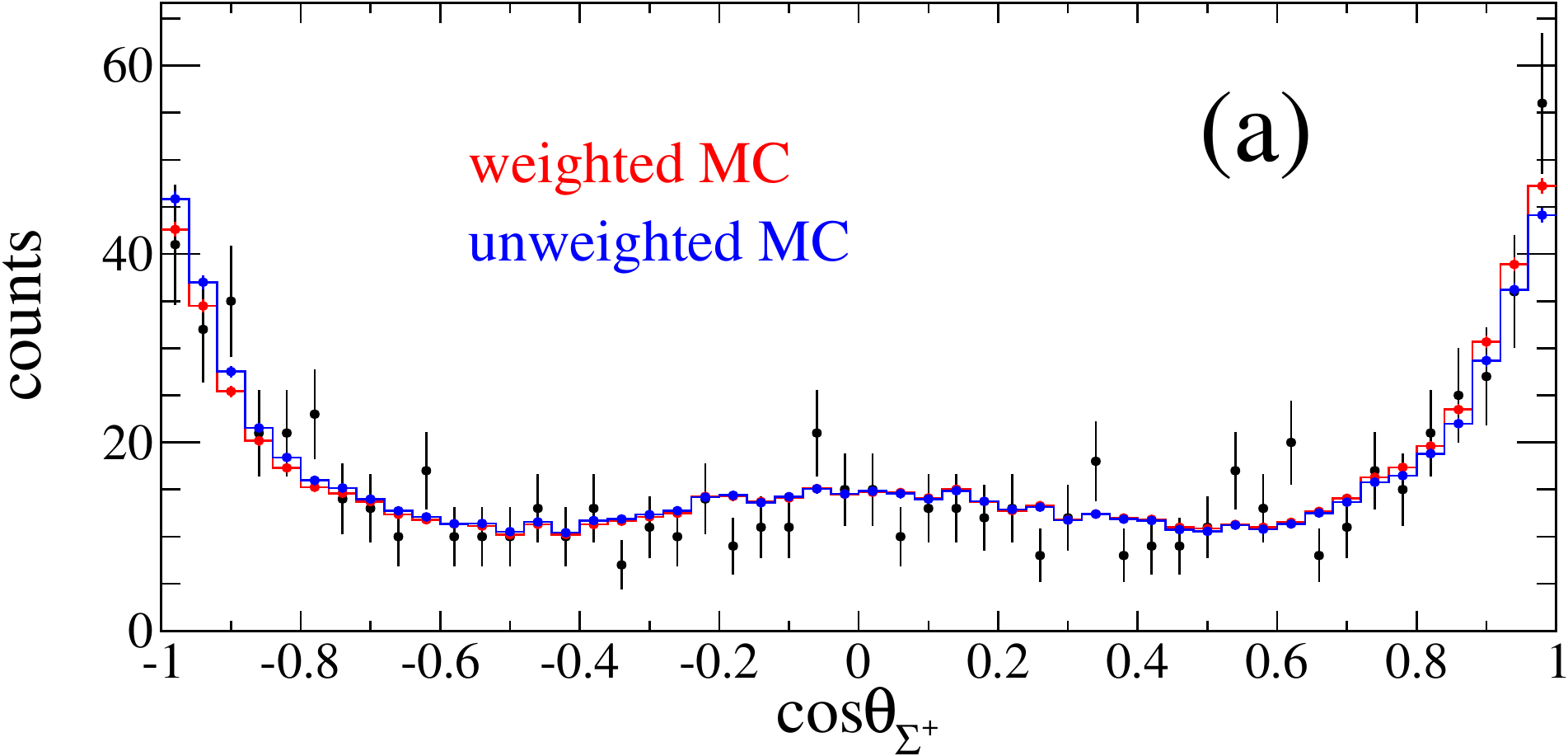}}
  \\
  \subfloat{\includegraphics[width=\linewidth]{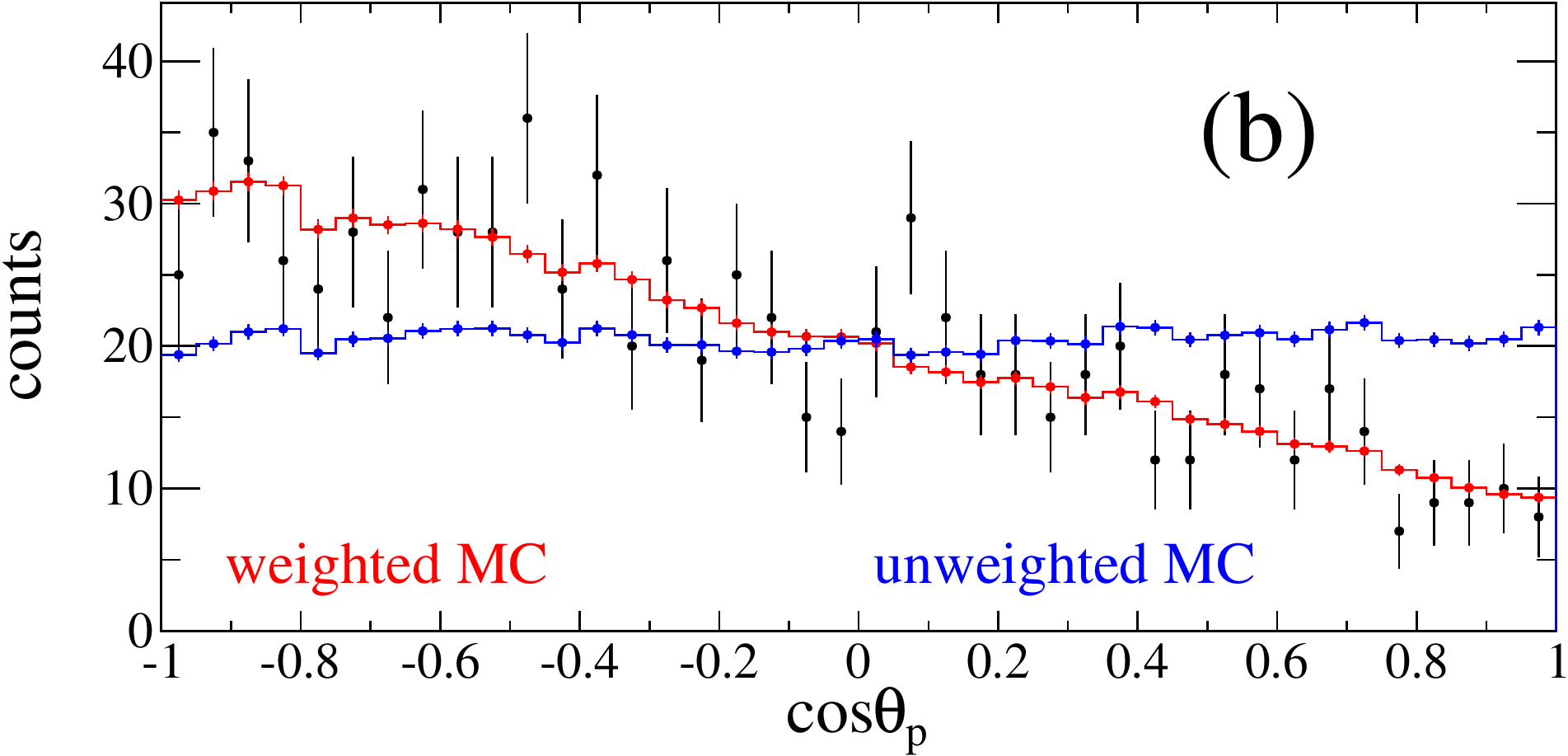}}
  \caption{(Color online) Distributions of the projections of (a)
  $\cos \theta_{\SigmaPlus}$ and (b) $\cos \theta_{\proton}$ for
    $2.65<W<2.75$~GeV and $0.70 < \costhetakp <
  0.80$. The black points are data, the blue histograms with points
  are the initial MC events without weighting, and the red histograms
  are the MC events weighted with the fit results using the
  $\frac{1}{2}^-$ hypothesis. Each of the MC histograms have been scaled to
  have the same area as the corresponding data histograms.  }
  \label{fig:SigmaPlusdist} 
\end{figure}

For each of the nine bins of $W$ and kaon angle, each spin-parity
hypothesis was tested with maximum likelihood fits to the data using a
MC simulation of the data that was matched to have the same
$\SigmaPlus\pim$ invariant mass distribution as the data but
generated without any angular correlations. The fit functions used
joint probability distributions of the $\SigmaPlus \pim$ angular decay
distribution and the $\proton \pizero$ weak decay distribution.  The
$\SigmaPlus \pim$ distribution was isotropic for spin $\frac{1}{2}$ and as given
in Eq.~\eqref{eq:distThreeHalf} for spin $\frac{3}{2}$. The $\proton
\pizero$ weak decay distribution used $I(\theta_{p}) \propto 1 +
\alpha_0 Q_z \cos \theta_{p}$, with the polarization $Q_z$ as a fit
parameter, and $\theta_p$ is the proton decay angle in the $\Sigma^+$
rest frame.  Figure~\ref{fig:SigmaPlusdist} shows sample distributions
of data and MC events for $\cos \theta_{\SigmaPlus}$ (our specific
$\cos\theta_Y$) and $\cos \theta_p$.
The nonflatness of the $\cos \theta_{\SigmaPlus}$ distribution
reflects the CLAS acceptance, which varies significantly depending not
only on $\cos \theta_{\SigmaPlus}$ and $\cos \theta_{\proton}$ but
also the azimuthal angles for each distribution.  We see from
Fig.~\ref{fig:SigmaPlusdist} that the unweighted MC that was generated
with isotropic distributions is able to reproduce the data well for
the $\cos \theta_{\SigmaPlus}$ distribution, lending support to the
spin $\frac{1}{2}$ hypothesis.  However, the $\cos \theta_{\proton}$ distribution
requires a reweighting of MC events with a polarization to match the
data, and it is this polarization of the data that allows a strong
discrimination among the different hypotheses.

The $\frac{3}{2}^{\pm}$ hypotheses were tested but showed no significant
deviation from an isotropic $\SigmaPlus \pim$ decay distribution, and
the parameter $p$ given in Eq.~\eqref{eq:distThreeHalf} was seen to be
consistent with $\frac{1}{2}$ (unpolarized).  For each separate
hypothesis, each of the MC events was assigned a weight according to
its fitted intensity. The resulting distribution was compared with the
data to calculate a $\chi^{2}$ probability. The $\chi^{2}$ probability
calculated for the $\cos \theta_{\proton}$ distribution had the most
discriminating power, and the $\frac{1}{2}^{-}$ case consistently had the
best $\chi^{2}$ probability. 
In our nine independent kinematic bins, the $\frac{1}{2}^{+}$ and
$\frac{3}{2}^{-}$ hypotheses are typically ruled out by $3 \sigma$ or more
from the $\chi^{2}$ probabilities and can be excluded. 
The three parameters describing the $\frac{3}{2}^{+}$ hypothesis can conspire to
exactly mimic the behavior of a $\frac{1}{2}^{-}$ state, so definitive exclusion
based on statistical tests is impossible.  Fits to the $\frac{3}{2}^{+}$ hypothesis
had worse $\chi^{2}$ probabilities in all energy bins, but we also excluded it
by assuming the simpler hypothesis with fewer parameters is correct.

\begin{figure}[h!t!p!b!]
  \includegraphics[width=\linewidth]{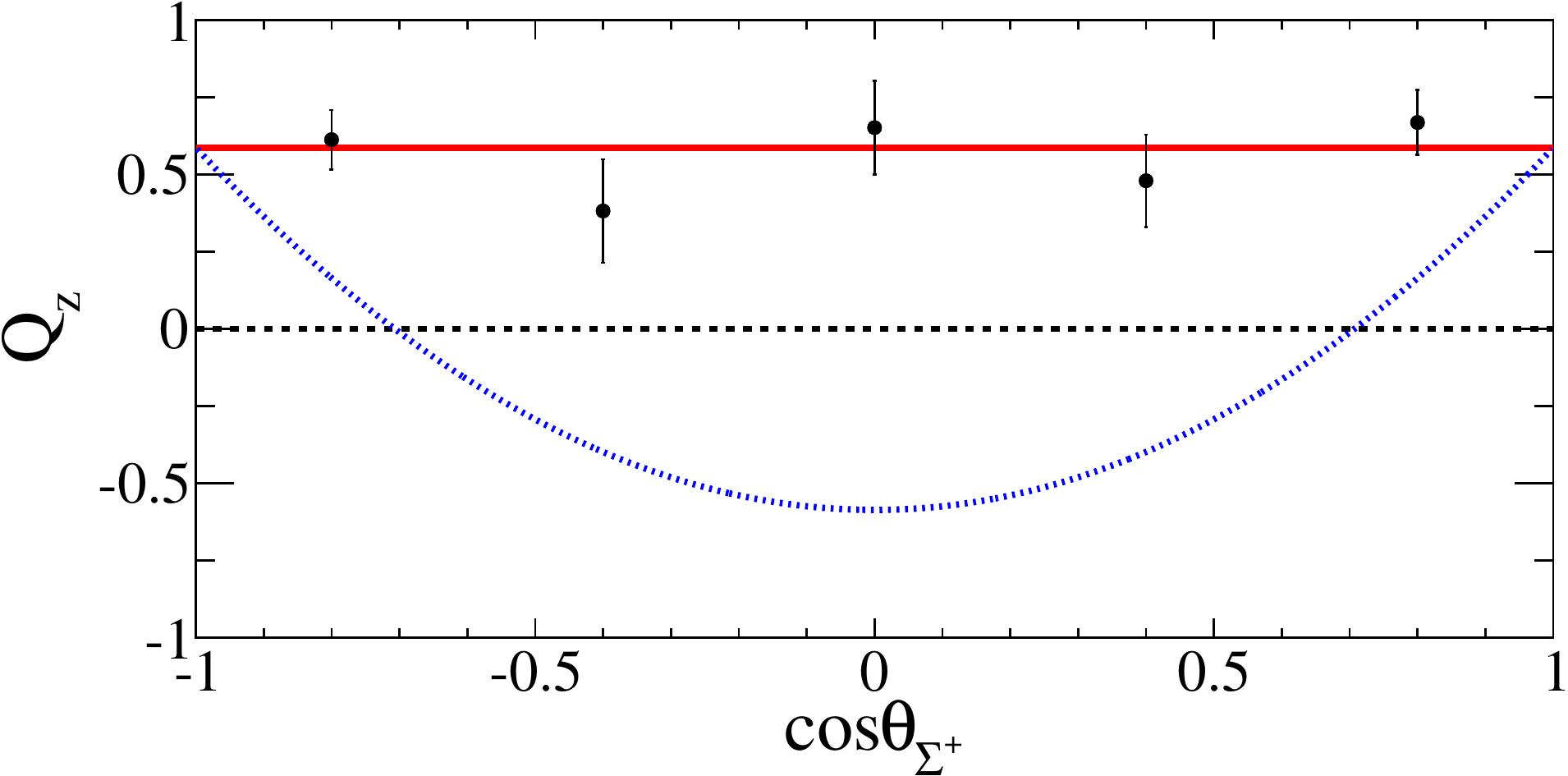}
  \caption{(Color online) Polarization $Q_{z}$ of $\SigmaPlus$ versus
  $\cos \theta_{\SigmaPlus}$ for
    $2.65<W<2.75$~GeV
  and $0.70 < \costhetakp < 0.80$. The average
  is shown as the red solid line. The dotted blue curve is
  the expectation
  for $P$-wave decay.
  }
  \label{fig:polAgainstSigmaAngle} 
\end{figure}

For the cases of spin-parity $\frac{1}{2}^{\pm}$, the two distinct behaviors of
the transferred polarization allow a simple visual
illustration. Independent fits were performed in separate bins of $\cos
\theta_{\SigmaPlus}$.  An example of the polarization $Q_z$ in the
$\hat{z}$ direction for one bin of $W$ and angle is shown in
Fig.~\ref{fig:polAgainstSigmaAngle}. As a function of $\cos
\theta_{\SigmaPlus}$, the polarization clearly does not change sign
between the extremes of $\cos \theta_{\SigmaPlus} = \pm 1$ and $\cos
\theta_{\Sigma} = 0$, as would be expected from Eq.~\eqref{eq:pwavePol}.
We can compute the probability of each hypothesis and while the
$\frac{1}{2}^{-}$ hypothesis consistently gives a good $\chi^{2}$
probability, the $\frac{1}{2}^{+}$ hypothesis is ruled out in its most
favorable bin by at least $3.6 \sigma$
and is typically ruled out by more than $5 \sigma$. 
With nine independent $W$ and angle bins,
the $\frac{1}{2}^{+}$ hypothesis is overwhelmingly ruled out.

The $\Sigma^+$ polarizations using all events in each kinematic bin
with the $\frac{1}{2}^{-}$ hypothesis are shown in Table~\ref{tab:pol_xy}.
Since in this situation the polarization that is measured through the
$\SigmaPlus$ is equivalent to the polarization of the $\LambdaOne$
itself, the $Q_z$ values in Table~\ref{tab:pol_xy} represent
measurements of the $\LambdaOne$ polarization.

\begin{table*}[h!t!]
  \setlength{\tabcolsep}{12pt}
  \centering
  \begin{tabular}{p{3.5cm}p{2.5cm}rrp{3.0cm}}
    \hline \hline
    $W$ (GeV) & \costhetakp    & \multicolumn{1}{c}{$Q_{x}$}            & \multicolumn{1}{c}{$Q_{y}$}            & $Q_{z} = \norm{\vec{P}}$ \\
    \hline
    $2.55$--$2.65$  &	$0.60$--$0.70$ & $ -0.06 \pm 0.06$  & $ -0.02 \pm 0.06$ & $0.38 \pm0.05$ \\
    $2.55$--$2.65$  &	$0.70$--$0.80$ & $ -0.04 \pm 0.05$  & $  0.01 \pm 0.05$ & $0.43 \pm0.05$ \\
    $2.55$--$2.65$  &	$0.80$--$0.88$ & $ -0.09 \pm 0.07$  & $  0.06 \pm 0.07$ & $0.47 \pm0.06$ \\
    $2.65$--$2.75$  &	$0.60$--$0.70$ & $ -0.09 \pm 0.07$  & $ -0.11 \pm 0.06$ & $0.36 \pm0.07$ \\
    $2.65$--$2.75$  &	$0.70$--$0.80$ & $  0.03 \pm 0.06$  & $  0.08 \pm 0.06$ & $0.59 \pm0.06$ \\
    $2.65$--$2.75$  &	$0.80$--$0.86$ & $  0.14 \pm 0.09$  & $ -0.01 \pm 0.09$ & $0.38 \pm0.08$ \\
    $2.75$--$2.85$  &	$0.60$--$0.70$ & $  0.05 \pm 0.08$  & $ -0.11 \pm 0.08$ & $0.40 \pm0.08$ \\
    $2.75$--$2.85$  &	$0.70$--$0.80$ & $ -0.02 \pm 0.07$  & $  0.08 \pm 0.07$ & $0.55 \pm0.06$ \\
    $2.75$--$2.85$  &	$0.80$--$0.86$ & $ -0.05 \pm 0.09$  & $ -0.13 \pm 0.10$ & $0.48 \pm0.09$ \\ \hline
    Full range      &	Full range     & $ -0.03 \pm 0.02$  & $  0.00 \pm 0.02$ & $0.45 \pm0.02$ \\ \hline \hline
  \end{tabular}

  \caption{Measured $\Sigma^+$ polarization components with
  statistical uncertainties in the $x$, $y$, and $z$ directions for
  each kinematic bin.  The $z$ direction is the out-of-plane
  production and polarization direction.  The $y$ direction is the
  initial-state proton axis, and the $x$ direction is
  $\hat{y}\times\hat{z}$.  } \label{tab:pol_xy}
\end{table*}


To ensure that the polarization we observe is not affected by the
$\Sigma^{0}(1385)$, the range of $\SigmaPlus \pim$ invariant mass was
changed to reduce this contribution. With a range of $1.40$--$1.48
~\gevcc$, we estimate the contribution of the $\Sigma^{0}(1385)$ as
$6\%$, and a change in polarization of $0.06$ was observed. Removal of
events with the $\kp \pim$ invariant mass within $\pm 1 \Gamma$ of the
$K^{\ast 0}$, where $\Gamma$ is the full width of the $K^{\ast 0}$,
gave a change of $0.02$ in the final result.

As a final check that the CLAS detector is able to measure decay
distributions without bias using this method, the polarization
components along the $x$ and $y$ directions were measured. If the
\LambdaOne{} is a state of $\frac{1}{2}^{-}$, then the polarization
components in the production plane should be zero.
Table~\ref{tab:pol_xy} shows that the measured components $Q_x$ and
$Q_y$ are mostly consistent with zero within the statistical
errors. Using these fits to estimate the systematic uncertainty of the
analysis in measuring these polarizations accurately, we take $\pm
0.03$ as an upper limit for this systematic uncertainty.
We add the upper
limit polarization components in the $x$ and $y$ directions, the
uncertainty due to varying the $\SigmaPlus \pim$ mass range, and the
uncertainty due to the $K^{\ast 0}$ removal in quadrature to obtain
the final systematic uncertainty of $\pm 0.07$.


To summarize, our analysis indicates that the spin-parity of the
\LambdaOne{} is fully consistent with $J^P=\frac{1}{2}^{-}$, while
the $\frac{1}{2}^{+}$ and $\frac{3}{2}^{-}$ combinations are strongly
disfavored.  The $\frac{3}{2}^{+}$ combination cannot, in
principle, be ruled out by statistical tests, but it did not lead to
better fits, and so it is rejected.  The
decay angular distribution is consistent with isotropy, which strongly
favors spin $\frac{1}{2}$, and under this assumption, a direct
measurement of the parity has been carried out for the first time. The
data strongly indicate negative parity, due to the unchanging
direction of the daughter $\Sigma^+$ polarization with respect to the
\LambdaOne{} polarization direction.  Thus, this first complete
experimental test of the $\frac{1}{2}^{-}$ hypothesis confirms most
long-held expectations.
As an additional outcome, the polarization of the
\LambdaOne{} in photoproduction has been measured to be $0.45 \pm
0.02 (\text{stat}) \pm 0.07 (\text{syst})$ in the forward kaon angle region for
$2.55 < W <2.85$~GeV.

\begin{acknowledgments}
We thank Professor R.~Kraemer for helpful early discussions.
We acknowledge the outstanding efforts of the staff of the Accelerator
and Physics Divisions at Jefferson Lab that made this experiment
possible. The work of the Medium Energy Physics group at Carnegie
Mellon University was supported by DOE Grant No. DE-FG02-87ER40315.  The
Southeastern Universities Research Association (SURA) operated the
Thomas Jefferson National Accelerator Facility for the United States
Department of Energy under Contract No. DE-AC05-84ER40150.  Further
support was provided by the National Science Foundation, the United
Kingdom's Science and Technology Facilities Council, and  the Italian 
Istituto Nazionale di Fisica Nucleare.
\end{acknowledgments}
\vfill

\bibliography{spinparity}

\end{document}

%% file: authors.tex
%
%
%
%
%
\newcommand*{\CMU}{Carnegie Mellon University, Pittsburgh, Pennsylvania 15213, USA}
\affiliation{\CMU}
\newcommand*{\ANL}{Argonne National Laboratory, Argonne, Illinois 60439, USA}
\affiliation{\ANL}
\newcommand*{\ASU}{Arizona State University, Tempe, Arizona 85287, USA}
\affiliation{\ASU}
\newcommand*{\CSUDH}{California State University, Dominguez Hills, Carson, CA 90747, USA}
\affiliation{\CSUDH}
\newcommand*{\CANISIUS}{Canisius College, Buffalo, NY 14208, USA}
\affiliation{\CANISIUS}
\newcommand*{\CUA}{Catholic University of America, Washington, D.C. 20064, USA}
\affiliation{\CUA}
\newcommand*{\SACLAY}{CEA, Centre de Saclay, Irfu/Service de Physique Nucl\'eaire, 91191 Gif-sur-Yvette, France}
\affiliation{\SACLAY}
\newcommand*{\UCONN}{University of Connecticut, Storrs, Connecticut 06269, USA}
\affiliation{\UCONN}
\newcommand*{\EDINBURGH}{Edinburgh University, Edinburgh EH9 3JZ, United Kingdom}
\affiliation{\EDINBURGH}
\newcommand*{\FU}{Fairfield University, Fairfield CT 06824, USA}
\affiliation{\FU}
\newcommand*{\FIU}{Florida International University, Miami, Florida 33199, USA}
\affiliation{\FIU}
\newcommand*{\FSU}{Florida State University, Tallahassee, Florida 32306, USA}
\affiliation{\FSU}
\newcommand*{\GWUI}{The George Washington University, Washington, DC 20052, USA}
\affiliation{\GWUI}
\newcommand*{\ISU}{Idaho State University, Pocatello, Idaho 83209, USA}
\affiliation{\ISU}
\newcommand*{\INFNFE}{INFN, Sezione di Ferrara, 44100 Ferrara, Italy}
\affiliation{\INFNFE}
\newcommand*{\INFNFR}{INFN, Laboratori Nazionali di Frascati, 00044 Frascati, Italy}
\affiliation{\INFNFR}
\newcommand*{\INFNGE}{INFN, Sezione di Genova, 16146 Genova, Italy}
\affiliation{\INFNGE}
\newcommand*{\INFNRO}{INFN, Sezione di Roma Tor Vergata, 00133 Rome, Italy}
\affiliation{\INFNRO}
\newcommand*{\ORSAY}{Institut de Physique Nucl\'eaire ORSAY, Orsay, France}
\affiliation{\ORSAY}
\newcommand*{\ITEP}{Institute of Theoretical and Experimental Physics, Moscow, 117259, Russia}
\affiliation{\ITEP}
\newcommand*{\JMU}{James Madison University, Harrisonburg, Virginia 22807, USA}
\affiliation{\JMU}
\newcommand*{\KNU}{Kyungpook National University, Daegu 702-701, Republic of Korea}
\affiliation{\KNU}
\newcommand*{\LPSC}{LPSC, Universite Joseph Fourier, CNRS/IN2P3, INPG, Grenoble, France}
\affiliation{\LPSC}
\newcommand*{\UNH}{University of New Hampshire, Durham, New Hampshire 03824, USA}
\affiliation{\UNH}
\newcommand*{\NSU}{Norfolk State University, Norfolk, Virginia 23504, USA}
\affiliation{\NSU}
\newcommand*{\OHIOU}{Ohio University, Athens, Ohio  45701, USA}
\affiliation{\OHIOU}
\newcommand*{\ODU}{Old Dominion University, Norfolk, Virginia 23529, USA}
\affiliation{\ODU}
\newcommand*{\URICH}{University of Richmond, Richmond, Virginia 23173, USA}
\affiliation{\URICH}
\newcommand*{\ROMAII}{Universita' di Roma Tor Vergata, 00133 Rome, Italy}
\affiliation{\ROMAII}
\newcommand*{\MSU}{Skobeltsyn Nuclear Physics Institute at Moscow State University, 119899 Moscow, Russia}
\affiliation{\MSU}
\newcommand*{\SCAROLINA}{University of South Carolina, Columbia, South Carolina 29208, USA}
\affiliation{\SCAROLINA}
\newcommand*{\JLAB}{Thomas Jefferson National Accelerator Facility, Newport News, Virginia 23606, USA}
\affiliation{\JLAB}
\newcommand*{\UTFSM}{Universidad T\'{e}cnica Federico Santa Mar\'{i}a, Casilla 110-V Valpara\'{i}so, Chile}
\affiliation{\UTFSM}
\newcommand*{\GLASGOW}{University of Glasgow, Glasgow G12 8QQ, United Kingdom}
\affiliation{\GLASGOW}
\newcommand*{\VIRGINIA}{University of Virginia, Charlottesville, Virginia 22901, USA}
\affiliation{\VIRGINIA}
\newcommand*{\WM}{College of William and Mary, Williamsburg, Virginia 23187, USA}
\affiliation{\WM}
\newcommand*{\YEREVAN}{Yerevan Physics Institute, 375036 Yerevan, Armenia}
\affiliation{\YEREVAN}
 
\newcommand*{\NOWUTFSM}{Universidad T\'{e}cnica Federico Santa Mar\'{i}a, Casilla 110-V Valpara\'{i}so, Chile}
\newcommand*{\NOWINFNFE}{INFN, Sezione di Ferrara, 44100 Ferrara, Italy}
\newcommand*{\NOWROMAII}{Universita' di Roma Tor Vergata, 00133 Rome Italy}
\newcommand*{\NOWINDIANA}{Indiana University, Bloomington, Indiana 47405, USA}
\newcommand*{\NOWSIENA}{Siena College, Loudonville, NY 12211, USA}
\newcommand*{\NOWWASHJEFF}{Washington \& Jefferson College, Washington, PA 15301, USA} 
\newcommand*{\NOWMIT}{MIT, Cambridge, MA 02139, USA} 


\author {K.~Moriya} 
\altaffiliation[Current address: ]{\NOWINDIANA}
\affiliation{\CMU}
\author {R.A.~Schumacher} 
\email[Contact: ]{schumacher@cmu.edu}
\affiliation{\CMU}
\author {M.~Aghasyan} 
\affiliation{\INFNFR}
\author {M.J.~Amaryan} 
\affiliation{\ODU}
\author {M.D.~Anderson} 
\affiliation{\GLASGOW}
\author {S.~Anefalos~Pereira} 
\affiliation{\INFNFR}
\author {J.~Ball} 
\affiliation{\SACLAY}
\author {N.A.~Baltzell} 
\affiliation{\ANL}
\affiliation{\SCAROLINA}
\author {M.~Battaglieri} 
\affiliation{\INFNGE}
\author {M.~Bellis} 
\altaffiliation[Current address: ]{\NOWSIENA}
\affiliation{\CMU}
% ...

\collaboration{CLAS Collaboration}
\noaffiliation